\documentclass[aps,prl,preprint,superscriptaddress]{revtex4-1}

\usepackage{soul}
\usepackage{color}
\usepackage{amsmath}
\usepackage{graphicx}
\usepackage{epstopdf}
\usepackage{mathtools}
\usepackage{url}
\usepackage{amsfonts}

\begin{document}
	\title{Experimental observation of \textit{PT} symmetry breaking near divergent exceptional points}
	\author{M. Sakhdari}
		\affiliation{Department of Electrical and Computer Engineering, University of Illinois at Chicago, Chicago, Illinois 60607, USA}
		
	\author{M. Hajizadegan}
	\affiliation{Department of Electrical and Computer Engineering, University of Illinois at Chicago, Chicago, Illinois 60607, USA}
	
	\author{Q. Zhong}
	\affiliation{Department of Physics and Henes Center for Quantum Phenomena, Michigan Technological University, Houghton, Michigan 49931, USA}
	
	\author{D.N. Christodoulides}
	\affiliation{College of Optics $\&$ Photonics-CREOL, University of Central Florida, Orlando, Florida 32816, USA}
	
	\author{R. El-Ganainy}
	\email[Corresponding author: ]{ganainy@mtu.edu}
	\affiliation{Department of Physics and Henes Center for Quantum Phenomena, Michigan Technological University, Houghton, Michigan 49931, USA}
	
	\author{P.-Y. Chen}
	\email[Corresponding author: ]{pychen@uic.edu}
	\affiliation{Department of Electrical and Computer Engineering, University of Illinois at Chicago, Chicago, Illinois 60607, USA}

\begin{abstract}
Standard exceptional points (EPs) are non-Hermitian degeneracies that occur in open systems. At an EP, the Taylor series expansion becomes singular and fails to converge---a feature that was exploited for several applications. Here, we theoretically introduce and experimentally demonstrate a new class of parity-time symmetric systems [implemented using radio frequency (rf) circuits] that combine EPs with another type of mathematical singularity associated with the poles of complex functions. These nearly divergent exceptional points  can exhibit an unprecedentedly large eigenvalue bifurcation beyond those obtained by standard EPs. Our results pave the way for building a new generation of telemetering and sensing devices with superior performance.
\end{abstract}
\maketitle

Spectral points that poss special features have been a subject of intense study recently. A well-known example of such points that are pertinent to periodic systems is the Van Hove singularity, at which the optical density of state does not vary smoothly as a function of frequency. First investigated in the context of lattice vibrations \cite{Van1953PR}, and later in photonic crystals \cite{Ibanescu2006PRL}, identifying these points has been proven useful in spectroscopy applications \cite{Dresselhaus2002C}. Another important class of spectral points are those associated with the eigenvalue degeneracy of Hermitian Hamiltonians [widely known as diabolic points (DPs)], which play an important role in the studies of molecular vibrations within the so called Born-Oppenheimer approximation. In the theory of band structures, a DP associated with a dispersionless band is also known as Dirac points \cite{Huang2011NM,Sepkhanov2007PRA} (since it also arises from the relativistic Dirac equation). While Dirac points are not associated with any topological protection, a close cousin, known as a Weyl point, further offers topological features \cite{Lu2013NP,Lu2014NP,Lu2015S}. Despite the fact that these mathematical constructions have been known for several decades, it was not until recently that physicists were able to experimentally probe them in the laboratory, especially in optical platforms where many-body interactions can be controlled at will.    

The aforementioned work focused mainly on Hermitian systems. Relaxing this condition to deal with effective non-Hermitian systems can result in even more exotic spectral features. More specifically, the non-Hermiticity of an effective Hamiltonian implies that its eigenstates do not need to be orthogonal. As a result, special degeneracies where both the eigenvalues and eigenfunctions become the same can occur at the so called exceptional points (EP) \cite{Heiss1990JPA,Magunov1999JPB,Heiss2004JPA,Heiss2012JPA,Rotter2003PRE,Muller2008JPA}. The interest in the peculiar behavior associated with EPs has exploded in the past years following the discovery of parity-time (\textit{PT}) symmetric Hamiltonians that exhibit real spectra \cite{Bender1998PRL}, and the introduction of this concept to classical wave dynamics for the first time \cite{Ruter2010NP,Makris2008PRL,Musslimani2008PRL,El-Ganainy2007OL}, which opened the door for a host of experimental studies in optics \cite{Peng2014S,Feng2014S,Hodaei2014S} and electronics \cite{Schindler2011PRA,Chitsazi2017PRL,Assawaworrarit2017N, Chen2018NE,Sakhdari2018IEEE,Hajizadegan2019IEEE,Chen2016PRAppl,Dong2019NE,Chen2019NE} as well as other platforms. Currently, several research groups are exploring the utility of non-Hermitian optics near EP to build miniaturized optical isolators \cite{Aleahmad2017SR,Ramezani2010PRA}, better lasers \cite{Feng2014S,Hodaei2014S,El-Ganainy2015PRA,Hokmabadi2019S,Gao2017O,Gu2016LPR,Liertzer2012PRL,Brandstetter2014NC,Peng2014S,El-Ganainy2014PRA,Sakhdari2018PRApp}, more responsive sensors \cite{Wiersig2014PRL,Wiersig2016PRA,Chen2017N,Hodaei2017N,Zhong2019PRL,Sakhdari2017NJP}, and nonlinear optical devices \cite{El-Ganainy2015OL,Zhong2016NJP} to mention just a few examples. For recent reviews, see \cite{El-Ganainy2018NP,Feng2017NP}. Additionally, enhanced wireless sensing with EPs is also attracting attention recently \cite{Assawaworrarit2017N, Chen2018NE,Sakhdari2018IEEE,Hajizadegan2019IEEE,Chen2016PRAppl,Dong2019NE,Chen2019NE}.

Despite this progress, all these activities focused only on one type of EPs having the form $\sqrt[n]{.}$. These represent branch point singularities at which the Taylor series expansion of the associated function fails to exist. However, apart from the discontinuity across the branch cut (with its intriguing implications for the encircling of EPs \cite{Doppler2016N,Xu2016N,Zhong2018NC}), the eigenvalues themselves (or equivalently, the associated multivalued function) remain bounded. 

In this Letter, we consider a rather unusual scenario where an EP coincides with (or occurs in the vicinity of) a divergent singularity. We show that, even though in practice physical systems cannot diverge, they can be locked near a divergent EP (DEP). Particularly, we show theoretically and demonstrate experimentally that the effect of a DEP on a nearby nondivergent EP can leave a clear fingerprint featured by giant enhancement of the eigenvalue splitting across the latter.

To this end, and before we describe our experimental results, let us consider a function of the form $f(x)=\frac{1}{\sqrt{1-x^2}}$. The function $f$ is real valued for $x<1$ and imaginary for $x>1$ with an EP located at $x=1$ which is also the same point where the function diverges, i.e. $f(1)=\infty$. As a result, in contrast to standard EPs where the splitting of the real part scales smoothly (for example as a square root function for second-order EPs), here, it diverges abruptly. If one could implement this system experimentally, it would be the ultimate sensor with infinite responsivity to any infinitesimal perturbation. Unfortunately, in practice, this is not possible. Many realistic physical effect (such as existence of a source at very high frequencies, stabilities and nonlinearities) will come into play to prevent such a response. Thus, it may seem that this concept is of little practical use. However, before we give up, let us consider a related function that has an additional degree of freedom: $F(x,y)=\frac{\sqrt{\alpha_1^2-y^2}}{\sqrt{\alpha_2^2-x^2}}$. Assuming $\alpha_1 \neq \alpha_2$, the function $F(x,y)$  will have a standard exceptional line at $y=\alpha_1$ and a divergent singularity at $x=\alpha_2$. If one can design a system that operates close enough to $x=\alpha_2$, the divergent point will be avoided, while at the same time its impact will be imprinted on the eigenvalue splitting across the exceptional line $y=\alpha_1$: the closer we get to $x=\alpha_2$, the larger the eigenvalue bifurcation. In this case, we call the EP $y=\alpha_1$ nearly divergent or NDEP (in the above example it is actually a line rather than a point, but this is irrelevant to the subsequent discussion). 

Having introduced the notion of DEPs and NDEPs theoretically, it is natural to inquire about the possibility of building a physical system that exhibits these spectral features. This is a challenging task given that optical systems (the most widely used platform for investigating non-Hermitian physics) do not naturally exhibit these spectral divergences, due to lack of lumped elements (in which the current does not vary, i.e., phase change or transition time is negligible). In this regard, radio frequency (rf) quasistatic resonators made of $RLC$ circuits [consisting of a resistor ($R$), an inductor ($L$), and a capacitor ($C$)] provide an advantage: in coupled \textit{PT} electronic systems (composed of two coupled $-RLC$ and $RLC$ resonators), the solution of the second-order differential equations arising from Kirchhoff current and voltage laws exhibit such a singularity \cite{Schindler2011PRA,Chen2018NE}. However, it occurs only for a perfect mutual coupling between the inductors, a condition that is impossible to achieve in practice. To complicate things further, it is not even easy to design a system that operates near this point. In practice, a nearly perfect inductive coupling requires a high permeability magnetic core and shielding plates, and any material or Eddy-current loss could decrease the coupling coefficient significantly \cite{Gan2006Thesis}.  

\begin{figure*}[!tb]
	\includegraphics[width=7in]{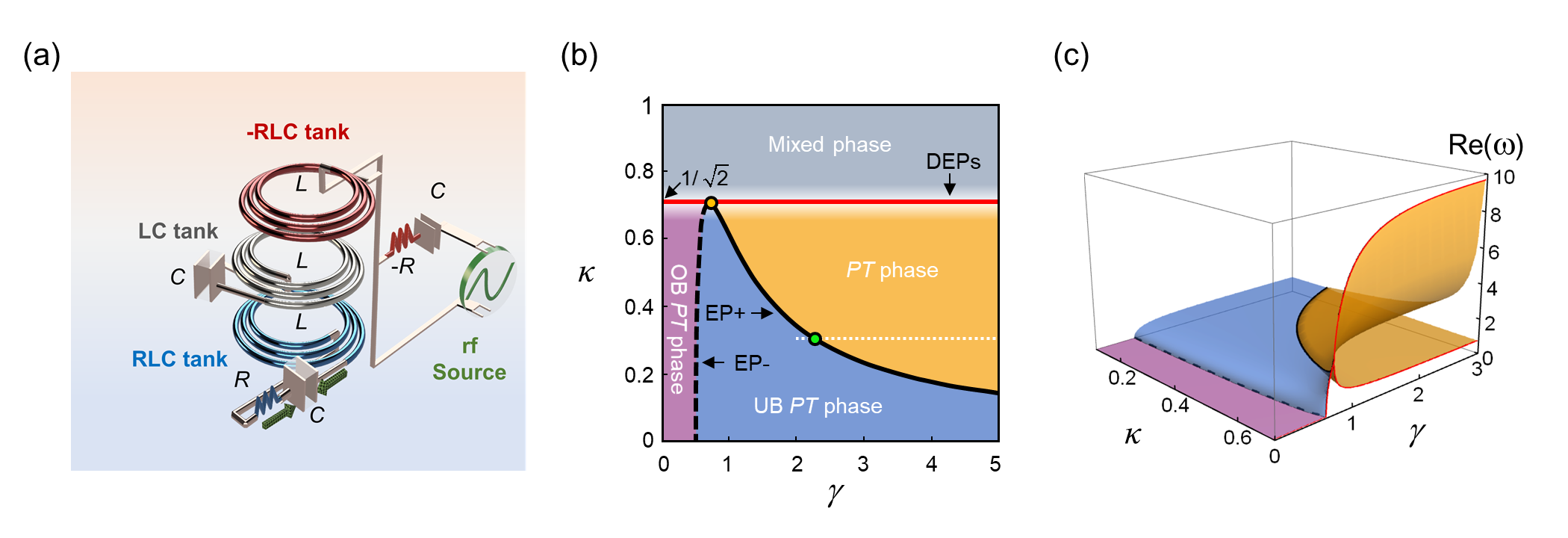}
	\caption{(a) A schematic of the three-elements \textit{PT}-symmetric electronic circuit proposed for implementing nearly divergent EPs. It consists of a $-RLC$ gain tank (top red), an $RLC$ loss tank (bottom blue) and a neutral element $L$C (center gray). The normalized coupling between the coils is $\kappa$ and the non-Hermitian parameter is $\gamma$ (see text for definition). (b) The phase diagram of this circuit in the $\kappa$--$\gamma$ plane. As discussed in the text, four different phases are identified: \textit{PT} symmetry, underdamped broken \textit{PT} symmetry (UB \textit{PT}), overdamped broken \textit{PT} symmetry (OB \textit{PT}) and a mixed phase that contains eigenstates in the \textit{PT} phase and others in the broken phase. The black solid and dashed lines, denoted by EP$\pm$, are exceptional lines that separate different phases. The solid red line consists of divergent EPs and separates the mixed phase from the rest of the domains. The white dashed line indicates the parameters used for the experiment as discussed later. (c) Bifurcation of real parts of the eigenvalues associated with (b). Note that as $\kappa \to \kappa_D=1/\sqrt{2}$, the splitting between the eigenfrequency becomes larger (theoretically diverges when $\kappa=\kappa_D$).   }
	\label{Fig-PhaseDiagram} 
\end{figure*}

In order to proceed, let us consider a coupled electronic circuit that consists of three stages representing gain, neutral, loss resonators, as shown in Fig. \ref{Fig-PhaseDiagram}(a). We will denote this circuit with $\text{C}_3$ (as opposed to the standard two-elements \textit{PT} circuit which we will denote $\text{C}_2$). One may think that adding a neutral element may lead to a standard higher order EP similar to the counterpart optical systems \cite{Demange2012JPA,Teimourpour2014PRA,Zhong2018}. However, this  is not the case. By applying Kirchhoff laws to the proposed circuit topology shown in Fig.  \ref{Fig-PhaseDiagram}(a). we can write an effective \textit{PT}-symmetric Hamiltonian for the system, $H_\text{eff}$, with the following eigenfrequencies [see Appendix A for more details]
\begin{equation} \label{Eq-evalues}
\omega_n=\pm 1, \pm \sqrt{\frac{2\gamma^2-1\pm \sqrt{1-4\gamma^2+8\gamma^4 \kappa^2}}{2\gamma^2(1-2\kappa^2)}},
\end{equation} 
where $\gamma=R^{-1} \sqrt{L/C}$ is the non-Hermitian parameter and $\kappa=M/L<1$ is  the normalized mutual coupling (here $M$ and $L$ are mutual and self inductances of the coils). By inspecting Eq. (\ref{Eq-evalues}),  it is clear that $\kappa=\kappa_D\equiv1/\sqrt{2}$ are the DEPs. For $\kappa < \kappa_D$, we can identify three different phases, separated by two exceptional lines described by the equations $\gamma_{\text{EP} \pm}=\sqrt{1 \pm \sqrt{1-2\kappa^2}}/(2\kappa)$, as shown as black solid or dashed lines in Fig. \ref{Fig-PhaseDiagram}(b). In the \textit{PT} phase given by $\gamma \in [\gamma_\text{EP+}, +\infty]$, all the eigenstates respect \textit{PT} symmetry with real eigenvalues. In the range $\gamma \in [\gamma_{\text{EP}-}, \gamma_\text{EP+}]$, the system exhibits an underdamped broken \textit{PT} (UB \textit{PT}) phase where the eigenvalues are complex conjugate. In the overdamped broken \textit{PT}  (OB \textit{PT}) phase with $\gamma \in [0,\gamma_{\text{EP}-}]$, the eigenvalues are purely imaginary. Mathematically, the last two regimes are separated by an exceptional line.  This feature arises due to charge conjugation symmetry of the Hamiltonian: $H=-H^*$ (see Appendix A for details). However, we note that, physically, the solutions related by this symmetry correspond to the same state.  For $\kappa > \kappa_D$, the eigenspectrum exhibits a mix between \textit{PT} states and broken \textit{PT} states. The boundary separating this mixed phase from the rest of the phase diagram is marked by a divergent exceptional line: a line made of DEPs, i.e., EPs that also coincide with pole singularities. Note that the three exceptional lines (black solid, black dashed and red lines) meet at one point given by $(\gamma,\kappa)=(1/\sqrt{2},1/\sqrt{2})$.  
It is important to emphasize that, before the system approaches this divergent regime, nonlinear effects dominated by the nonlinearity of active circuit element (which is used to implement the negative resistance as discussed in Appendix B) will come into play to regulate the circuit behavior. Thus, the important question is, can one at least engineer the system to operate close enough to these DEPs such that they have significant impact on the spectral features? Figure \ref{Fig-PhaseDiagram}(c) plots the linear spectrum associated with Eq. (\ref{Eq-evalues}). It shows that close to the  DEPs, the eigenvalue's bifurcation (which corresponds to frequency splitting between two resonant frequencies, not to their amplitudes) becomes dramatic---a feature that can be utilized to build next generation ultra-responsive \textit{PT} sensors beyond the current state of the art. In theory, similar behavior can be also traced in the conventional two-elements \textit{PT} symmetric systems studied in \cite{Schindler2011PRA}. In practice, the divergent exceptional line in the latter occurs for $\kappa=1$---a condition that is impossible to achieve in experiment as it implies that perfect mutual coupling between the inductors, i.e. equal values for the mutual and self inductances. Thus the main merit of the three-elements circuit presented here is to bring these singularities to an experimentally accessible domain. Importantly, we note that the above results do not have analogs in optical systems. In fact, an optical \textit{PT} trimer that consists of neutral element sandwiched between gain and loss sites will demonstrate a very different behavior by possessing a third order exceptional point \cite{Hodaei2017N,Demange2012JPA,Teimourpour2014PRA,Zhong2018}.

\begin{figure}[!tb]
	\includegraphics[width=4in]{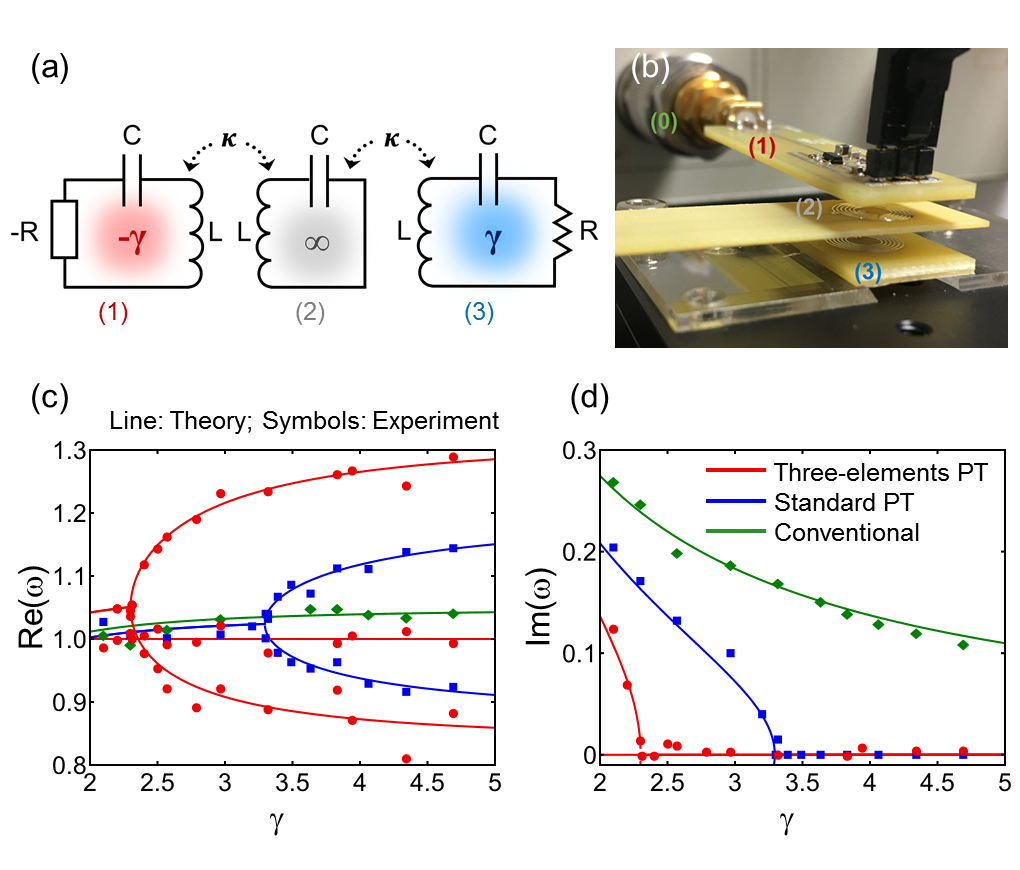}
	\caption{Circuit schematic (a) and picture of the on-board circuit implementation (b) of the proposed three-elements rf network. (c) and (d) plot the real and imaginary eigenfrequencies as extracted from the rf reflection measurements (see Appendix C for details). From (c), it is clear that the frequency splitting in the proposed system (red lines and dots) is larger than that of the corresponding standard \textit{PT} system having a coupling $\kappa=0.3$ (blue lines and dots), as well as that associated with conventional telemetry system based on non-\textit{PT} inductive coupling geometry (see Fig. 6 in Appendix B) which is shown in green lines and dots \cite{Chen2018NE}.}
	\label{Fig-Exp1} 
\end{figure}

In order to demonstrate the advantage of the proposed circuit topology [Fig. \ref{Fig-PhaseDiagram}(a)] in providing indirect access to the DEPs with potential telemetric sensing applications, we have built a prototype using onboard circuit technology (see Appendix B for details). The circuit consists of a tunable $RLC$ tank that mimics a wireless capacitive sensor \cite{Chen2018NE}. This pseudosensor consists of a variable capacitor, connected in series to a planar spiral inductor and a resistor (which accounts for the effective resistance of the sensor), such that its equivalent circuit is identical to that of a realistic wireless sensor. The information provided by the sensor is then read by an $-RLC$ tank connected to the vector network analyzer for measuring the reflection spectrum. Unlike standard \textit{PT}-symmetric telemetric systems where the sensor and reader tanks are directly coupled \cite{Chen2018NE}, the current system is constructed by inserting a neutral $LC$ tank between the $-RLC$ and $RLC$ oscillators as shown in Fig. \ref{Fig-PhaseDiagram}(a). In both the $-RLC$ and $RLC$ resonators, the inductance of microstrip coils is $L = 330 \ \text{nH}$ and the absolute value of resistance
$|-R|=R= 50 \ \Omega$. In order to emulate behaviors of a wireless capacitive sensor, the capacitance $C$ of tank circuits is tuned from 30 to 220 pF (SMA CER $\pm$0.05 pF). This, in turn, varies the non-Hermiticity parameter of the system, $\gamma \propto 1/\sqrt{C}$ which is the relevant parameter for real-life wireless capacitive sensing applications \cite{Chen2018NE}.  A schematic diagram and a picture of the implemented circuit are shown in Fig. \ref{Fig-Exp1}(a) and \ref{Fig-Exp1}(b), respectively. For comparison, we have also fabricated a standard (two-element) \textit{PT} circuit. In both structures, the normalized coupling coefficient was engineered to be $\kappa=0.3$. While this value is relatively weak, it still favors the three-element circuit in terms of operation near the DEP (corresponding to $\kappa \approx 0.7$) as compared to the standard two-element circuit having DEP at $\kappa=1$.

\begin{figure}[!tb]
	\includegraphics[width=4in]{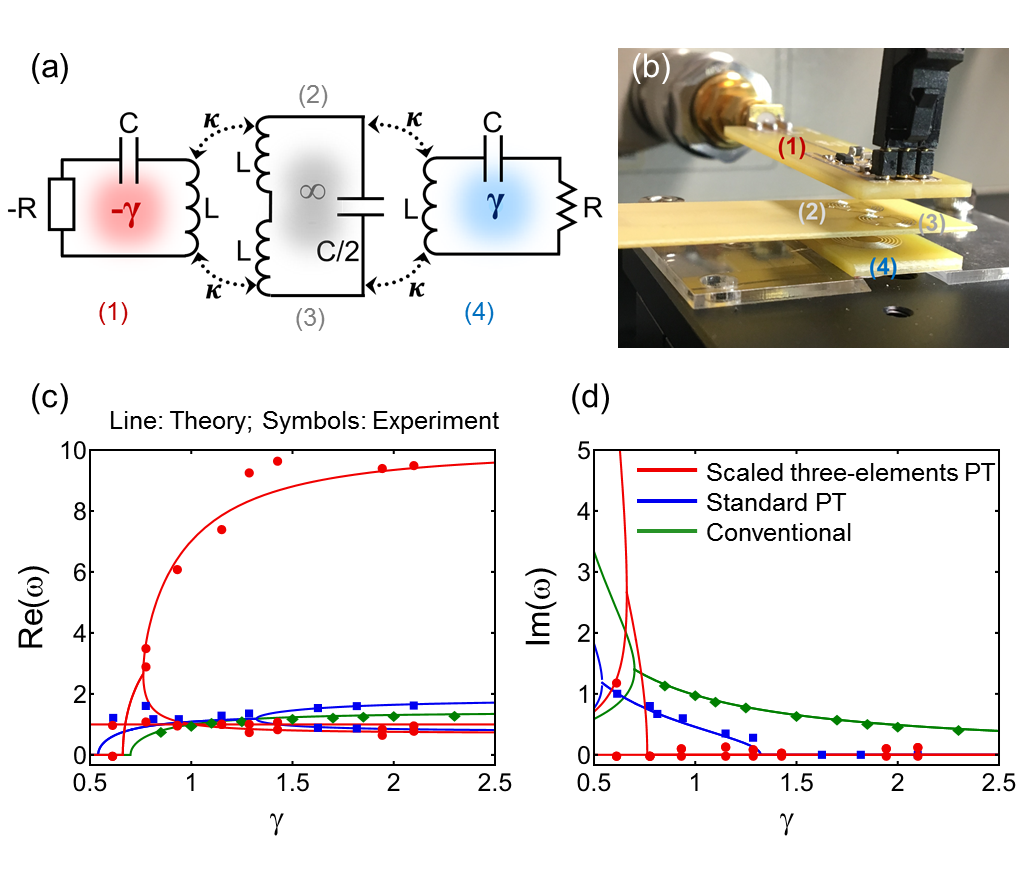}
	\caption{Schematic of the scaled three-element circuit (a) and picture of its on-board circuit implementation (b). (c) and (d) plot the real and imaginary eigenfrequencies, varying a function of non-Hermiticity parameter $\gamma$, for the dual-links three-stages \textit{PT}-symmetric telemetric system in (a) with $\kappa=0.495$  (red circles), the standard \textit{PT}-symmetric telemetric system with $\kappa=0.7$  (blue squares), and the conventional one using a micro-coil reader with $\kappa=0.7$ (green diamonds). A five-fold enhancement in the bifurcation compared to standard \textit{PT} circuit is observed. For comparison, we also present the results for conventional telemetry system (green line and dots) \cite{Chen2018NE}. For completeness, we also plot the constant eigenfrequency (horizontal red line) associated with the solution in Eq. (\ref{Eq-evalues}).}
	\label{Fig-Exp2} 
\end{figure}

Figures \ref{Fig-Exp1}(c) and \ref{Fig-Exp1}(d) plot the theoretical (solid lines) and experimental (dots) values of complex eigenfrequencies as a function of the non-Hermitian parameter $\gamma$ for the proposed three-element circuit. The experimental results here span the range indicated by the white dashed line in Fig. \ref{Fig-PhaseDiagram}(b), i.e., they trace the transition from the UB \textit{PT} phase to the \textit{PT} phase across the EP marked by the green point in the figure. For comparison, we also present  the results for the standard two-element \textit{PT} circuit  on the same figure. First, we find a good agreement between theoretical predictions and experimental data. Second, it is clear that the three-element system demonstrates giant frequency splitting (red data points) compared with the standard one (blue dots). Finally, we also note that the location of the EP in the proposed three-element system is down shifted compared with the standard circuit, which is in agreement with theory.

Encouraged by these results, we have also explored the related system shown in Fig. \ref{Fig-Exp2}(a).  Here, the neutral oscillator has the same resonant frequency as before but with its inductor and capacitor scaled according to  $2 L$ and a  $C/2$. Furthermore, we consider the coupling topology shown in  Fig. \ref{Fig-Exp2}(a). By following the similar analysis to that shown in Appendix A, one can show that the new frequency splitting will be enhanced because $\kappa$ in Eq. (\ref{Eq-evalues}) is replaced by $\sqrt{2}\kappa$. In this case, the divergent exceptional line is located at $\kappa_D=0.5$. In other words, this modified circuit requires reduced normalized coupling to bring the system closer to the DEP, which in turn, leads to enhanced eigenvalue bifurcation. Figure \ref{Fig-Exp2}(b) depicts the fabricated circuit with the modified neutral circuit. The theoretical and experimental data for the spectral bifurcation are plotted in Figs. \ref{Fig-Exp2}(c) and \ref{Fig-Exp2}(d). Again for comparison, we also plot the data for the standard two-element \textit{PT} circuit. As evidenced by the plots, we observe a gigantic enhancement of the frequency bifurcations, almost 5 times more than in the previous case.

\begin{figure}[!tb]
	\includegraphics[width=4in]{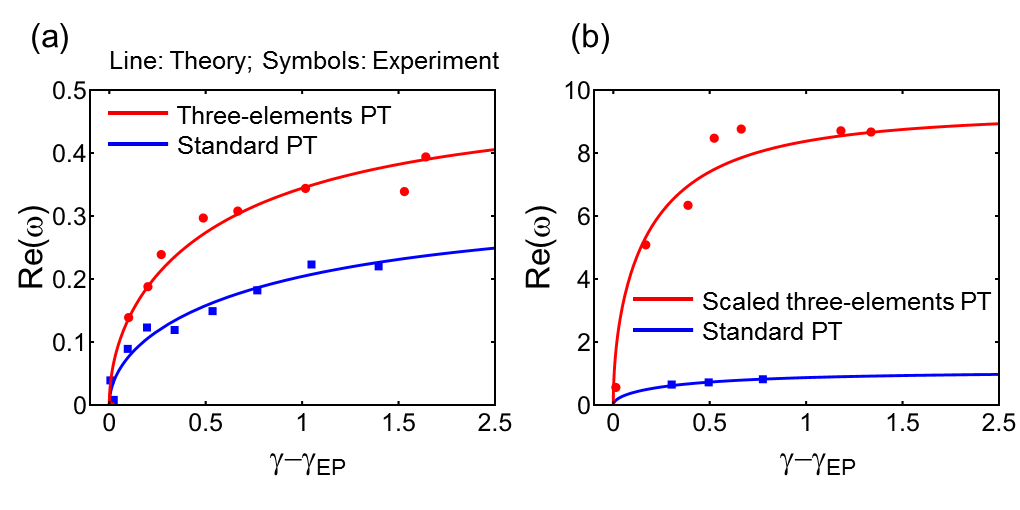}
	\caption{Plots of the frequency splitting as a function of $\Delta \gamma=\gamma-\gamma_\text{EP}$ for both experimental setups shown in Fig. \ref{Fig-Exp1} and \ref{Fig-Exp2}. Note that the scaled \textit{PT} circuit offers a clear advantage as measured by larger splitting.} 
	\label{Fig-FrequencySplitting} 
\end{figure}

To further facilitate the comparison between the proposed circuits with respect to each others as well as to the standard \textit{PT} circuit, we also plot the frequency splitting extracted from Figs. \ref{Fig-Exp1} and \ref{Fig-Exp2} as a function of $\Delta \gamma=\gamma-\gamma_\text{EP}$. As can be observed from Figs. \ref{Fig-FrequencySplitting}(a) and \ref{Fig-FrequencySplitting}(b), the scaled \textit{PT} circuit, being closer to the DEP, offers a clear advantage as measured by larger splitting.

In conclusion, we have introduced the notion of divergent exceptional points and showed how they can be indirectly accessed by using three-element \textit{PT}-symmetric electronic circuits made of gain-neutral-loss resonators. We have tested our predictions experimentally and demonstrated that, indeed, the eigenfrequency bifurcation close to divergent exceptional points can be boosted as a result of the interplay between the square root splitting of second order EPs and the giant multiplication factor associated with DEPs. We envision that such new non-Hermitian electronic systems, when applied to wireless probing and telemetering, will enable a superior sensing capability. This work can be also extended to other microwave, millimeter-wave and terahertz wireless systems.   

\begin{acknowledgments}
R.E. is supported by the Army Research Office (ARO) Grant No. W911NF-17-1-0481, and the National Science Foundation (NSF) Grant No. ECCS 1807552. P.-Y.C. is supported by NSF Grant No. ECCS 1914420: CAREER, and UIC DPI Cycle 2 Seed Funding Program.  

M.S., M.H. and Q.Z. contributed equally to this manuscript.
\end{acknowledgments}

\appendix
\subsection{Appendix A. Effective PT symmetric Hamiltonian for three coupled oscillators}
The PT symmetric circuit presented in Fig. 1(a) of the main text consists of $-$RLC, LC and RLC tanks. By applying the Kirchhoff’s circuit laws as expressed using the Liouvillian formalism \cite{Schindler2011PRA} to this system, we obtain:
\begin{equation}\label{Eq:Hamiltonian}
	\begin{aligned}
		\mathcal{L} \mathbf{\Psi} &=\frac{d \mathbf{\Psi}}{d \tau}, \\
		\mathcal{L} &=\begin{bmatrix}
			0 & 0 & 0 & 1 & 0 & 0 \\
			0 & 0 & 0 & 0 & 1 & 0 \\
			0 & 0 & 0 & 0 & 0 & 1 \\
			-\dfrac{1-\kappa^2}{1-2\kappa^2} & \dfrac{\kappa}{1-2\kappa^2} & -\dfrac{\kappa^2}{1-2\kappa^2} & \dfrac{1}{\gamma}\dfrac{1-\kappa^2}{1-2\kappa^2} & 0 & -\dfrac{1}{\gamma}\dfrac{\kappa^2}{1-2\kappa^2} \\
			\dfrac{\kappa}{1-2\kappa^2} & -\dfrac{1}{1-2\kappa^2} & \dfrac{\kappa}{1-2\kappa^2} & -\dfrac{1}{\gamma}\dfrac{1-\kappa^2}{1-2\kappa^2} & 0 & \dfrac{1}{\gamma}\dfrac{\kappa}{1-2\kappa^2} \\
			-\dfrac{\kappa^2}{1-2\kappa^2} & \dfrac{\kappa}{1-2\kappa^2} & -\dfrac{1-\kappa^2}{1-2\kappa^2} & \dfrac{1}{\gamma}\dfrac{1-\kappa^2}{1-2\kappa^2} & 0 & -\dfrac{1}{\gamma}\dfrac{1-\kappa^2}{1-2\kappa^2} 
		\end{bmatrix}
	\end{aligned},
\end{equation}
where $\mathbf{\Psi}\equiv(q_1,q_2,q_3,\dot{q}_1,\dot{q}_2,\dot{q}_3)^T$, $q_1$, $q_2$ and $q_3$ are charges stored on the capacitor in the $-$RLC, LC, and RLC tanks, $\gamma=R^{-1}\sqrt{L/C}$ is the non-Hermiticity parameter (note that positive and negative  respectively play the role of gain and loss), $\kappa=M/L$ is the normalized coupling strength ($0 \leq \kappa \leq 1$) and $M$ is the mutual inductance between two neighboring electronic oscillators. Additionally, $\tau \equiv \omega_0 t$, $\omega_0=1/\sqrt{LC}$ is the natural frequency of the isolated neutral LC tank, and all frequencies are measured in units of $\omega_0$.   From Eq. (\ref{Eq:Hamiltonian}), we find that the Liouvillian expression remains the same under the combined parity $\mathcal{P}$ ($q_1 \leftrightarrow q_3$) and time-reversal $\mathcal{T}$ ($t \rightarrow -t$) transformation. The above equation can be also cast in the Hamiltonian formalism by using $H_\text{eff}=i \mathcal{L}$. Note that the effective Hamiltonian $H_\text{eff}$ is non-Hermitian (i.e. $H_\text{eff}^\dagger \neq  H_\text{eff}$ ) and respects PT symmetry with $[\mathcal{PT},H_\text{eff}]=0$, where:
\begin{subequations}
	\begin{align}
		\mathcal{P}&=
		\begin{bmatrix}
			\mathbf{J} & 0 \\
			0 & \mathbf{J}
		\end{bmatrix}, \\
		\mathcal{T}&=\mathcal{K}
		\begin{bmatrix}
			\mathbf{I} & 0 \\
			0 & -\mathbf{I}
		\end{bmatrix}
	\end{align}
\end{subequations}
Here, $\mathbf{J}$ is the $3 \times 3$ an anti-diagonal matrix with unit entries, $\mathbf{I}$ is the $3 \times 3$ identity matrix, and $\mathcal{K}$ the antilinear complex conjugation operator. The eigenfrequencies of the above system, as expressed in the bases $e^{i \omega \tau}$  can be derived from the secular equation $H_\text{eff}-\omega_n \mathbf{I}=0$  with $n=1,2,...,6$,  which in turn gives Eq. (1) in the main text.

\begin{figure}[!t]
	\includegraphics[width=3in]{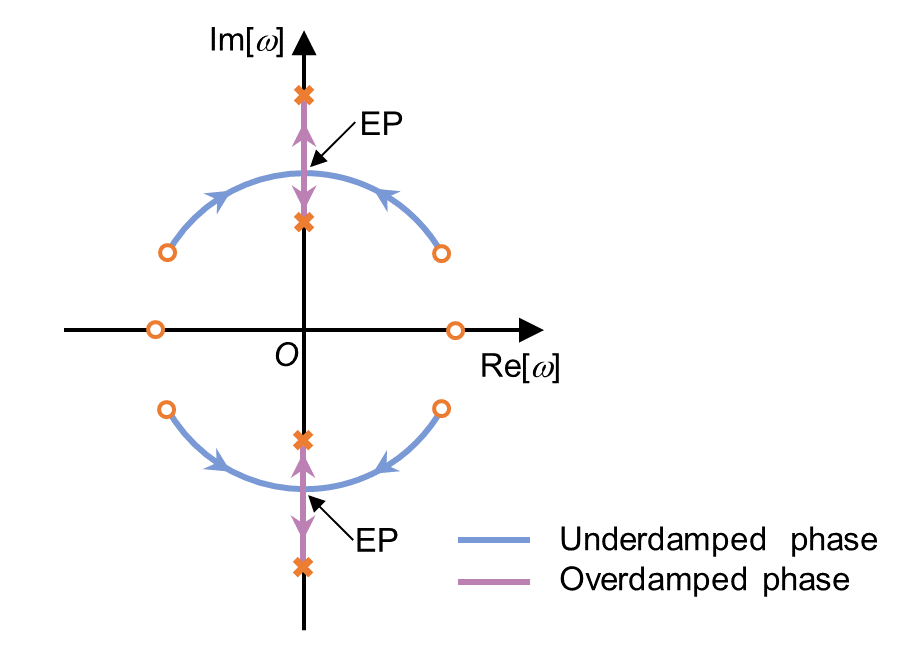}
	\caption{Trajectories of the eigenvalues associated with the Hamiltonian $H_\text{eff}$  for $\kappa=0.3$ and $\gamma \in [0.48,1]$ with $\gamma=1$ corresponding to the starting points. The PT symmetry of $H_\text{eff}$ guarantees that the complex conjugate of each eigenvalue is also an eigenvalue. On the other hand, the charge conjugation symmetry dictates that negative complex conjugate of each eigenvalue is also an eigenvalue. As a result, when the parameter $\gamma$ is varied in the example considered here, the eigenvalues to the left/right of the imaginary axis move towards the imaginary axis as shown until each pair meet at an EP before they bifurcate again on the imaginary axis. The open circles represent the initial eigenvalues while the crosses represent the final points on that particular trajectory. The underdamped and overdamped phases are marked by different colors that correspond to Fig. 1(b) of the main text. Importantly, we note that the presence of EPs here is a mathematical feature of the model. From a physical perspective, the eigenstates that correspond to the eigenvalues and their negative complex conjugate represent the same physical state.}
	\label{Fig-Phase_Transition} 
\end{figure}

Note that $H_\text{eff}=i \mathcal{L}$ is a pure imaginary matrix given that both $\gamma$ and $\kappa$ are real. Therefore, the Hamiltonian $H_\text{eff}$ posses a charge conjugation symmetry (also known as particle-hole symmetry) \cite{Ge2017PRA,Malzard2015PRL} , i.e. $H_\text{eff}=-H_\text{eff}^*$.  Together with PT symmetry, this dictates that the eigenvalues are either complex conjugates with a plus or minus sign prefactor. Thus, in addition to the conventional PT to BPT phase transition, the system exhibits also another phase transition in the BPT domain, from an underdamped phase with complex eigenvalues to an overdamped phase with pure imaginary eigenvalues. Despite being in the BPT phase, the underdamped/overdamped domains are separated by EPs (see Fig. \ref{Fig-Phase_Transition} for an intuitive proof).

\subsection*{Appendix B. RF reader designs and wireless measurement setups}
\begin{figure}[!t]
	\includegraphics[width=\linewidth]{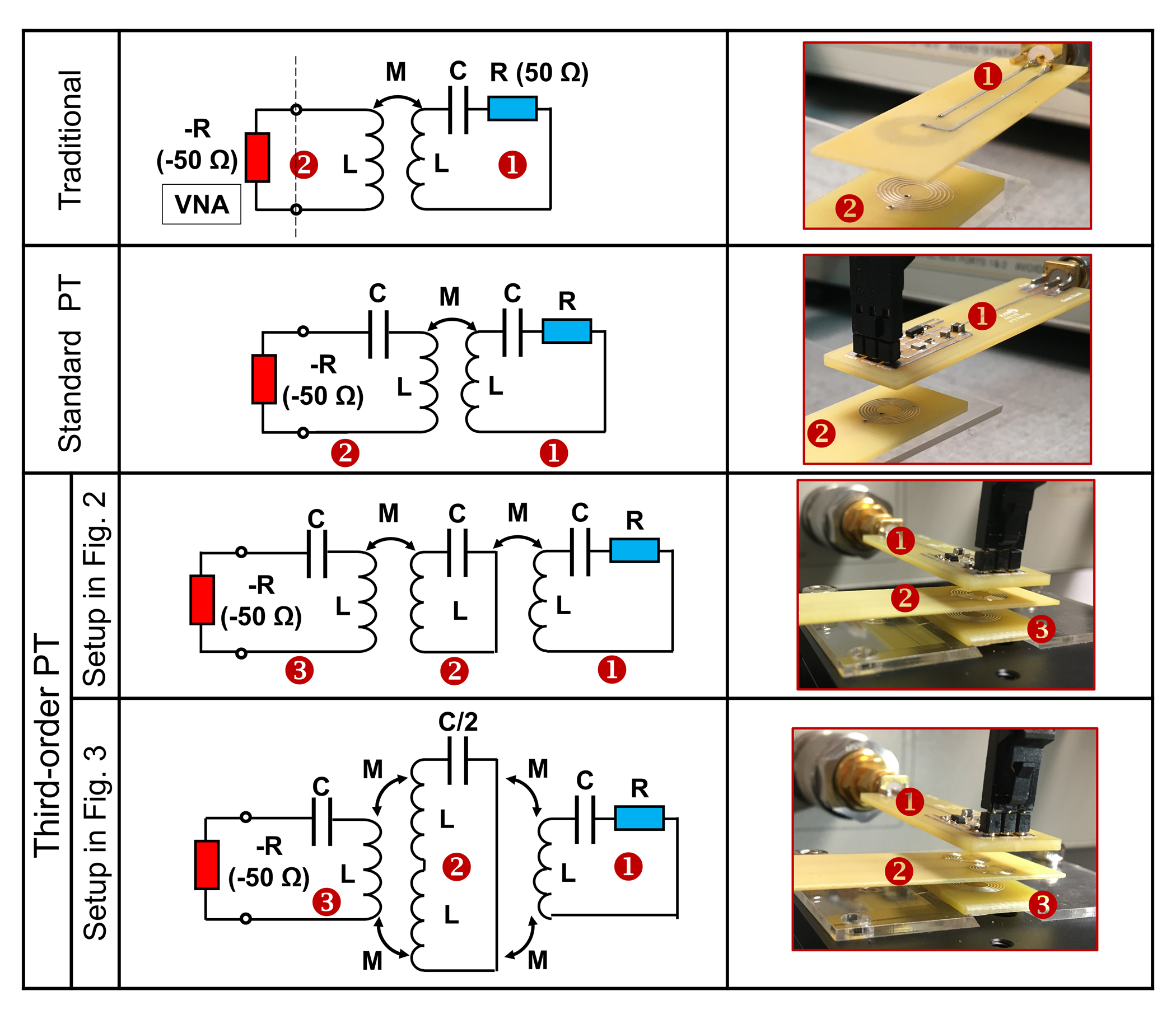}
	\caption{Schematics and experimental setups for traditional, standard PT-symmetric, and third-order PT-symmetric telemetry system. }
	\label{FigS2} 
\end{figure}

In our measurements, the reader (–RLC tank) was connected to the vector network analyzer or VNA (Agilent E5061B; $Z_0 = 50\ \Omega$), and the intermediator (LC tank) and the pseudo-sensor (RLC tank) were fixed on the XYZ linear translation stage. In the $-$RLC tank, the equivalent negative resistance can be realized with a negative impedance converter, such as a Colpitts-type circuit with a positive feedback. In addition, while a resistor accounts for power dissipation, a negative resistor means a source of energy in the closed-loop analysis. Hence, a RF signal generator with source impedance $Z_0$ (e.g., a VNA with a wideband, phase-locked loop (PLL)-based frequency synthesizer that feeds the reader) is described by a negative impedance $-Z_0$. Figure \ref{FigS2} presents schematic diagrams and experimental setups for the conventional, standard PT-symmetric, and third-order PT-symmetric telemetry systems. Traditional sensor telemetry systems usually employ a passive loop antenna for wireless readout (via inductive coupling). In the one-port measurement, the information is encoded in the reflection coefficient. For example, if the physical quantity of interest varies the capacitance of a passive wireless capacitive sensor (equivalent to a RLC oscillator), it can be characterized by tracking the resonance frequency shift. Under steady-state condition (i.e., charges and displacement currents in Eq. (2) have time-harmonic variations), zero reflection can be obtained at eigenfrequencies (resonant frequencies). For the standard PT-symmetric system, the reflection coefficient can be derived as:
\begin{equation}\label{Eq:Gamma2}
	\Gamma=\frac{\prod\limits_{n=1}^4 (\omega-\omega_n)}{\dfrac{2 \omega [\omega+i \gamma (\omega^2-1) ]}{\eta \gamma^2 (\kappa^2-1)}+\prod\limits_{n=1}^4 (\omega-\omega_n)},
\end{equation}
where $\eta=R/Z_0$  and $\omega_n$  is the $n$-th eigenfrequency of the system. For the third-order PT-symmetric system (Fig. 1(a) in the main text), the reflection coefficient is: 
\begin{equation}\label{Eq:Gamma3}
	\Gamma=\frac{\prod\limits_{n=1}^6 (\omega-\omega_n)}{\dfrac{2 \omega [\omega(\omega^2-1)+i \gamma (1-2\omega^2-(\kappa^2-1)\omega^4) ]}{\eta \gamma^2 (2\kappa^2-1)}+\prod\limits_{n=1}^6 (\omega-\omega_n)}.
\end{equation}
From Eqs. (\ref{Eq:Gamma2})-(\ref{Eq:Gamma3}), reflection dips are obtained at $\omega_n$  (in the unit of $\omega_0$). In the exact PT-symmetric phase ($\omega_n \in \mathbb{R}$), zero reflection can be achieved when $\omega=\omega_n$. The effect of zero reflection at resonance frequencies can be attributed to conjugate matching, for which power extracted from the external source (-R) is equal to that dissipated in R and reactive powers stored in L and C tend to cancel each other out.

\subsection*{Appendix C. Magnitude of the reflection coefficient versus frequency}

\begin{figure}[!ht]
	\includegraphics[width=5in]{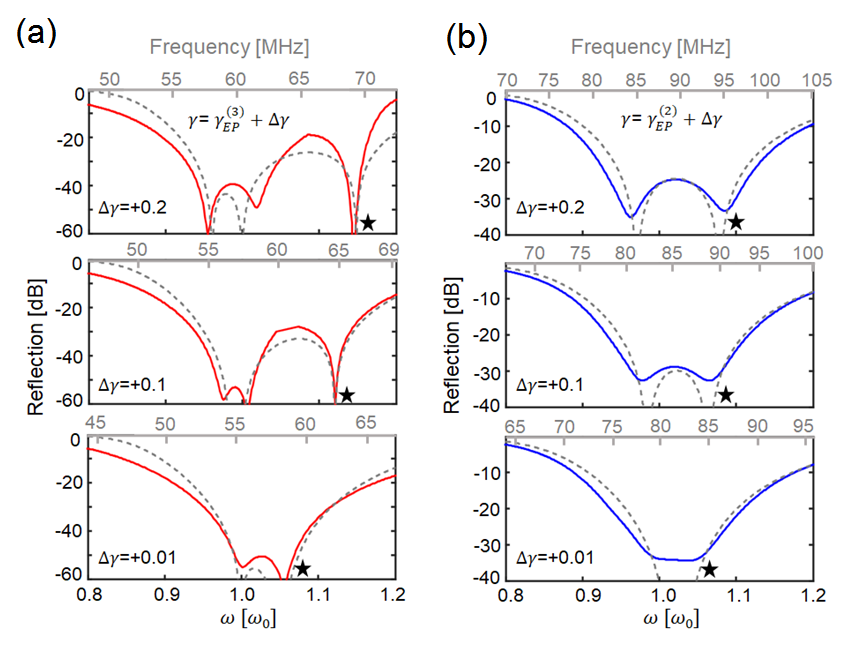}
	\caption{Magnitude of the reflection coefficient versus frequency associated with the experimental setup shown in Fig.1 in the main text, for (a) the three-elements PT-symmetric system  and, (b) the standard PT-symmetric system. Both systems are controlled to operate around the exceptional point (EP), under the coupling strength $\kappa=0.3$  Solid and dashed lines denote experimental and theoretical results, respectively. }
	\label{FigS3} 
\end{figure}

\break

\begin{figure}[!ht]
	\includegraphics[width=5in]{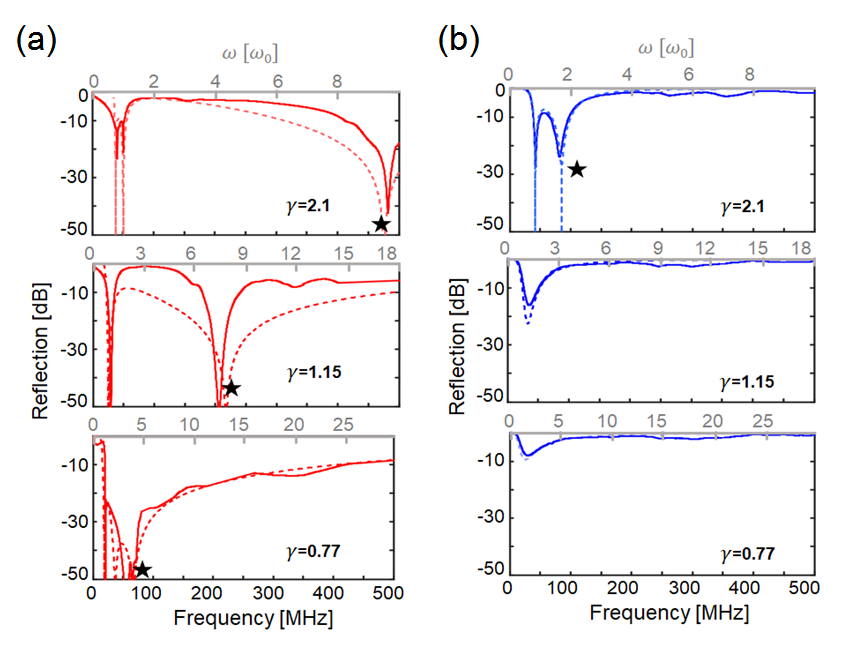}
	\caption{Magnitude of the reflection coefficient versus frequency for the four-links third-stages PT-symmetric telemetric system, in comparison with the standard PT-symmetric telemetric system; here, the non-Hermiticity paramaters $\gamma$ is varied from 0.77 to 2.1, and coupling strength $\kappa$ is fixed to 0.495. Note that if $\gamma< 1.3$, the standard PT-symmetric system ( $\gamma_\text{EP} = 1.3$ ) operates in the broken phase, which exhibits a weak resonance with a low quality factor. Solid and dashed lines denote experimetnal and theoretical results, respectively. }
	\label{FigS4} 
\end{figure}

\subsection*{Appendix D. Measurement precision}
In our experiment, the reflection data was collected using a vector network analyzer (VNA), series Agilent E5061B . In these devices, the dynamic accuracy of the receiver consists of compression, noise, and crosstalk. At low signal levels, the VNA receiver's measured power is composed of the signal power applied to the input of the VNA’s receiver and is added to the noise power, any residual crosstalk power, and the error in the measurement. When measuring a high-level signal (full reflect or transmission through a short transmission line or lumped-element circuits in this work). The measurement was performed by acquiring a minimum of 10 sweeps of data from the desired parameter, in linear magnitude and phase, after a trace-math normalization. At each frequency point, the population-based standard deviation is computed. At –60 dBm signal power (near exceptional points where $|S_{11}|\sim 0$), the noise and residual crosstalk contributions are 0.000609 dB peak with a 1 kHz IFBW (Intermediate Frequency Band Width) setting. At –60 dBm and below, the signal power level, the noise, and residual crosstalk contributions become more significant relative to other sources of measurement errors (see \cite{VNA} for more details). However, in our measurement, the standard deviation is below 0.001 dB, especially in the PT-symmetric phase. Therefore, electronic noises and errors sourced from the measurement equipment (VNA) do not cause any visible effect on the results presented in all figures.

\bibliography{Reference}

\end{document}